\documentclass[prx, aps, amsfonts, amssymb, amsmath, twocolumn, notitlepage, nofootinbib, twoside, showkeys, 10pt]{revtex4-2}
\usepackage[english]{babel}
\usepackage[utf8]{inputenc}
\usepackage[colorinlistoftodos, color=green!40, prependcaption]{todonotes}
\usepackage{amsthm}
\usepackage{mathtools}
\usepackage{physics}
\usepackage{xcolor}
\usepackage{graphicx}
\usepackage[left=23mm,right=13mm,top=35mm,columnsep=15pt]{geometry} 
\usepackage{adjustbox}
\usepackage{placeins}
\usepackage[T1]{fontenc}
\usepackage{lipsum}
\usepackage{csquotes}
\usepackage{bm}
\usepackage{placeins}
\usepackage[T1]{fontenc}
\usepackage{lipsum}
\usepackage{csquotes}
\usepackage[pdftex, pdftitle={Article}, pdfauthor={Author}, colorlinks=true,linkcolor=blue,citecolor=blue,urlcolor=blue,filecolor=magenta]{hyperref}
\usepackage{comment}
\usepackage{nameref}
\usepackage{url}
\usepackage[babel]{microtype}
\usepackage{pdfpages}

\makeatletter
\AtBeginDocument{\let\LS@rot\@undefined}
\makeatother

\def\supplementfilename{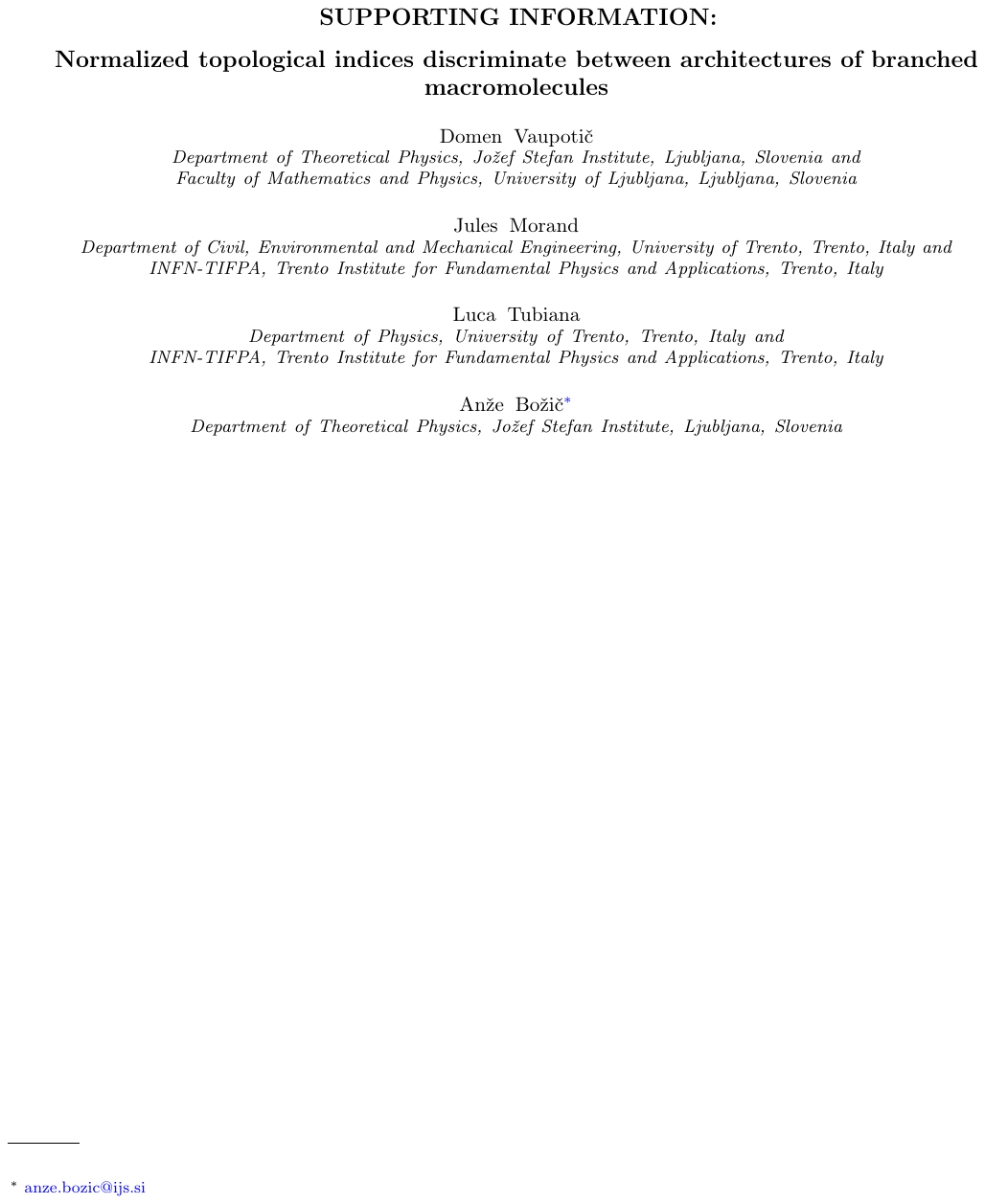}

\pdfximage{\supplementfilename}
\def\numbersupplementpages{\the\pdflastximagepages}

\usepackage{xr-hyper}

\makeatletter

\newcommand*{\addFileDependency}[1]{
\typeout{(#1)}
\@addtofilelist{#1}
\IfFileExists{#1}{}{\typeout{No file #1.}}
}\makeatother

\newcommand{\ABC}{\mathrm{ABC}}
\newcommand{\TI}{\mathrm{TI}}
\newcommand{\cdf}{\operatorname{cdf}}
\newcommand{\pdf}{\operatorname{pdf}}

\begin{document}

\title{Normalized topological indices discriminate between architectures of branched macromolecules}

\author{Domen Vaupoti\v{c}}
\affiliation{Department of Theoretical Physics, Jo\v{z}ef Stefan Institute, Ljubljana, Slovenia}
\affiliation{Faculty of Mathematics and Physics, University of Ljubljana, Ljubljana, Slovenia}
\author{Jules Morand}
\affiliation{Department of Civil, Environmental and Mechanical Engineering, University of Trento, 
Trento, Italy}
\affiliation{INFN-TIFPA, Trento Institute for Fundamental Physics and Applications, 
Trento, Italy}
\author{Luca Tubiana}
\affiliation{Department of Physics, University of Trento, 
Trento, Italy}
\affiliation{INFN-TIFPA, Trento Institute for Fundamental Physics and Applications, 
Trento, Italy}
\author{An\v{z}e Bo\v{z}i\v{c}}
\email{anze.bozic@ijs.si}
\affiliation{Department of Theoretical Physics, Jo\v{z}ef Stefan Institute, Ljubljana, Slovenia}

\begin{abstract}
Branching architecture characterizes numerous systems, ranging from synthetic (hyper)branched polymers and biomolecules such as lignin, amylopectin, and nucleic acids to tracheal and neuronal networks. Its ubiquity reflects the many favourable properties that arise because of it. For instance, branched macromolecules are spatially compact and have a high surface functionality, which impacts their phase characteristics and self-assembly behaviour, among others. The relationship between branching and physical properties has been studied by mapping macromolecules to mathematical trees whose architecture can be characterized using topological indices. These indices, however, do not allow for a comparison of macromolecules that map to trees of different size, be it due to different mapping procedures or differences in their molecular weight. To alleviate this, we introduce a novel normalization of topological indices using estimates of their probability density functions. We construct a phase space using two normalized topological indices, which enables a robust discrimination between different architectures of branched macromolecules. We demonstrate the necessity of such a phase space on two practical applications, one being ribonucleic acid (RNA) molecules with various branching topologies and the other different methods of coarse-graining branched macromolecules. Our approach can be applied to any type of branched molecules and extended as needed to other topological indices, making it useful across a wide range of fields where branched molecules play an important role, including polymer physics, green chemistry, bioengineering, biotechnology, and medicine.
\end{abstract}

\keywords{Branched macromolecules; topological indices; RNA branching; coarse-graining methods}

\maketitle

\section{Introduction}

Molecular topology permeates different systems and scales~\cite{tubiana2024topology}. Aside from knots and links, an important aspect of topology is the propensity of different macromolecules to assume {\em highly branched architectures}~\cite{qu2021macromolecular,bonchev1995topological}. Branching imbues macromolecules with favourable attributes such as high surface functionality, globular conformation, and high solubility~\cite{cook2020branched,kobayashi2023branching}, and consequently influences many of their properties, including phase characteristics, self-assembly behaviour, and encapsulation capability~\cite{zhu2011influence,adroher2023effect,jiang2015hyperbranched}.

Branched macromolecules exhibit a powerful interplay between their structure, properties, and function~\cite{seo2020beauty}, and it is thus no wonder that they occur and are utilized in numerous contexts: {\em Synthetic} branched, dendritic, and hyperbranched polymers and their derivatives have a rich history and a wide range of applications~\cite{voit2009hyperbranched,carlmark2013dendritic,jiang2015hyperbranched,cook2020branched}. Branching furthermore occurs naturally in different {\em biomolecules}, including amylopectin~\cite{thompson2000non}, ubiquitin~\cite{kolla2022assembly}, and lignin~\cite{balakshin2020spruce,wang2022ligningraphs,ratnaweera2015impact}, as well as in both deoxyribonucleic acid (DNA)~\cite{seeman1998nucleic,dong2020dna,mai2016single} and ribonucleic acid (RNA)~\cite{wiedemann2022rnaloops,vaupotivc2023scaling,vaupotivc2023viral}. Branching is an important phenomenon also at large scales, as demonstrated by branched tissues found in various organisms~\cite{goodwin2020branching,hannezo2017unifying}, such as the tracheal systems of insects~\cite{hayashi2018development,centanin2010tracheal}, mammalian lung epithelium and salivary and mammary glands~\cite{goodwin2020branching}, and the diverse neuronal systems of different species~\cite{markram2004interneurons,kanari2018topological}.

An attractive feature of both synthetic and naturally-occurring branched molecules is that their branching architecture can be controlled and interfered with~\cite{takata2018topology,ma2021highly,Guan1999,guan2002control}, which can be achieved either during their synthesis or, in the case of biopolymers, by genetic perturbations~\cite{ralph2019lignin}, modifications of branching enzymes~\cite{li2019starch,li2016progress}, and other mechanisms. Lignin, the second most abundant biological macromolecule after cellulose, can assume various branched and cross-linked topologies~\cite{balakshin2020spruce,wang2022ligningraphs,ratnaweera2015impact} and as such represents a molecule with a vast potential for the generation of renewable substitutes for synthetic materials~\cite{ralph2019lignin,zhang2022comparison}. Design and synthesis of various branched polymers has been utilized in order to increase lithium transport in them~\cite{hao2021lithium}, and branching has also been proposed to be the mechanism by which to overcome the limitations of linear polymers relevant in waste management and crude oil depletion~\cite{corneillie2015pla,pan2023progress}. Hyperbranched polymers and other branched molecules have furthermore potential biomedical applications~\cite{Saadati2021,newland2015prospects,dong2020dna}, including biomedical imaging, biosensors, tissue engineering, and drug or gene delivery systems.

The branching architecture of macromolecules can be described using topological parameters which ignore their three-dimensional conformation~\cite{rouvray1986predicting,rouvray2002topology,shimokawa2019topology}. Some of the parameters used to characterize branching include the number of termini and junctions~\cite{shimokawa2019topology}, the degree of branching~\cite{voit2009hyperbranched}, and in particular {\em topological indices}, which have been designed to cover a vast range of topological characteristics of branched macromolecules and relate them to their physicochemical properties~\cite{austel1983topological,gutman2013degree,stankevich1988topological}. The first topological index was introduced in 1947 by Wiener~\cite{wiener1947structural}, followed by eight decades of research and introduction of hundreds of different new indices~\cite{todeschini2008handbook}. They have proven to be particularly useful for predicting properties of molecules in quantitative structure-property/activity relationship (QSPR/QSAR)~\cite{dearden2017use}. However, not all topological quantities yield predictions of similar quality for different types of molecules and in general do not allow for a straightforward comparison of macromolecules of different sizes. This can happen either when considering molecules of the same type but different molecular weight (e.g., different RNAs), or when comparing different molecular species which might therefore be mapped to trees in different ways despite having a similar molecular weight.

In this work, we introduce a novel normalization of topological indices based on estimating their probability density functions. We choose two normalized topological indices to construct a phase space that enables robust discrimination between the topologies of molecules of different types and sizes. The necessity of our approach is illustrated with two practical examples. First, we use the phase space to distinguish between topologies of RNA secondary structures with different degrees of branching and to show how different tree representations of RNA influence the resulting topology. Second, we compare two methods of coarse-graining the structures of branched macromolecules and use the phase space of normalized topological indices to highlight the differences between the two methods. Taken together, our work establishes a phase space based on normalized topological indices that allows for a well-defined way of discriminating between branching architectures of different macromolecules as well as of methods of coarse-graining them.

\begin{figure}[!t]
    \centering
    \includegraphics[scale=1]{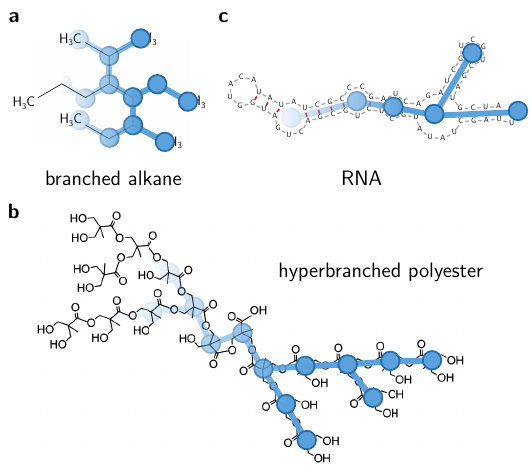}
    \caption{Mapping branched macromolecules to mathematical trees. Shown are the chemical structures of {\bf (a)} branched alkane and {\bf (b)} hyperbranched polyester as well as {\bf (c)} the secondary structure of an RNA molecule. Each panel illustrates how the branching structure of the macromolecule is mapped to nodes and edges of a corresponding tree. Panel (b) adapted from Ref.~\cite{carlmark2013dendritic} under CC BY 3.0 licence.}
    \label{fig1}
\end{figure}


\section{Phase space of normalized topological indices}

\subsection{Tree representation of branched macromolecules}

Branching is characteristic of macromolecules across different systems and scales (Fig.~\ref{fig1}). To compare their topologies, one can map their structure onto a mathematical tree $T$ with $N$ edges (the tree {\em size}), with $V(T)$ denoting the set of its nodes and $E(T)$ denoting the set of its edges. This mapping is generally not unique and depends on the macromolecule~\cite{shimokawa2019topology}. For instance, a molecular (chemical) graph has atoms as nodes and edges that represent chemical bonds (Fig.~\ref{fig1}a). Due to the nature of carbon bonding, chemical graphs have a maximum branching degree of $4$ regardless of their size. For polymer structures, a typical tree representation is one where a node is a collection of atoms at a polymer branch point and an edge is a linear structure formed by a chain of bonds between two such branch points (Fig.~\ref{fig1}b). A more complex case, RNA is a linear polymer whose constituent nucleotides interact and base-pair with one another and give rise to a complex and highly branched secondary structure. This structure is often mapped to a tree by associating the base-paired (double-stranded) regions with edges and unpaired (single-stranded) regions with nodes (Fig.~\ref{fig1}c). As we will see later on, such a coarse-grained fashion of mapping macromolecules to trees can lead to RNA trees with significantly larger branching degrees compared to chemical graphs.

\subsection{Constructing the phase space of topological indices}

Once a branched macromolecule is mapped to a mathematical tree, its architecture can be studied using methods from graph theory. This includes a wide variety of topological indices designed to describe the topology of trees and, more generally, graphs, with applications ranging from the description of chemical properties of molecules to connectivity of networks~\cite{austel1983topological,gutman2013degree,stankevich1988topological}. Broadly speaking, we can divide topological indices into path-, degree-, centrality-, spectrum-, and information-based indices depending on the main graph property used in their definition. For simplicity, we focus here on $18$ commonly used topological indices; their definitions are given in Supporting Information (SI). Since most topological indices were designed to be applied to graphs in general---which include cycles---they tend to be highly correlated when used on trees (Fig.~S1 in SI). {\em Individual} indices are also highly degenerate and do not discriminate well between different tree topologies~\cite{konstantinova2003discriminating}. Therefore, our first aim is to select a {\em pair} of topological indices which are correlated as little as possible so that we can use their phase space for comparison of tree topologies.

We find that out of the $18$ topological indices we examined, one of the least correlated pairs (Fig.~S1) is formed by the second-order network coherence $H_2$ and the atom bond connectivity index $\ABC$. The second-order network coherence is spectrum-based and defined as
\begin{equation}
\label{eq:h2}
    H_2=\frac{1}{2N}\sum_{i=2}^N\frac{1}{\lambda_i^2},
\end{equation}
where $\lambda_i$ are the eigenvalues of the Laplacian matrix, $\mathcal{L}=\mathcal{D}-\mathcal{A}$, and $\mathcal{D}$ and $\mathcal{A}$ are the degree matrix and the adjacency matrix of the tree (or, more generally, graph), respectively. The degree matrix is a diagonal matrix of node degrees $d(i)$, $\mathcal{D} = \mathrm{diag}(d(i))$, while the adjacency matrix is a $(0,1)$-matrix where $\mathcal{A}_{ij}=1$ if and only if nodes $i$ and $j$ are connected by an edge. Second order network coherence has mainly been used in networks to characterize the variance of fluctuations in consensus systems with additive noise~\cite{patterson2014consensus}. The atom bond connectivity index is degree-based and defined as
\begin{equation}
\label{eq:abc}
    \ABC=\sum_{(i,j)\in E(T)}\left[\frac{d(i)+d(j)-2}{d(i)d(j)}\right]^{1/2}.
\end{equation}
It has first been used as a molecular structure descriptor to study different properties of alkanes~\cite{estrada1998atom,estrada2008atom}.

\begin{figure*}[!htp]
    \centering
    \includegraphics[width=\linewidth]{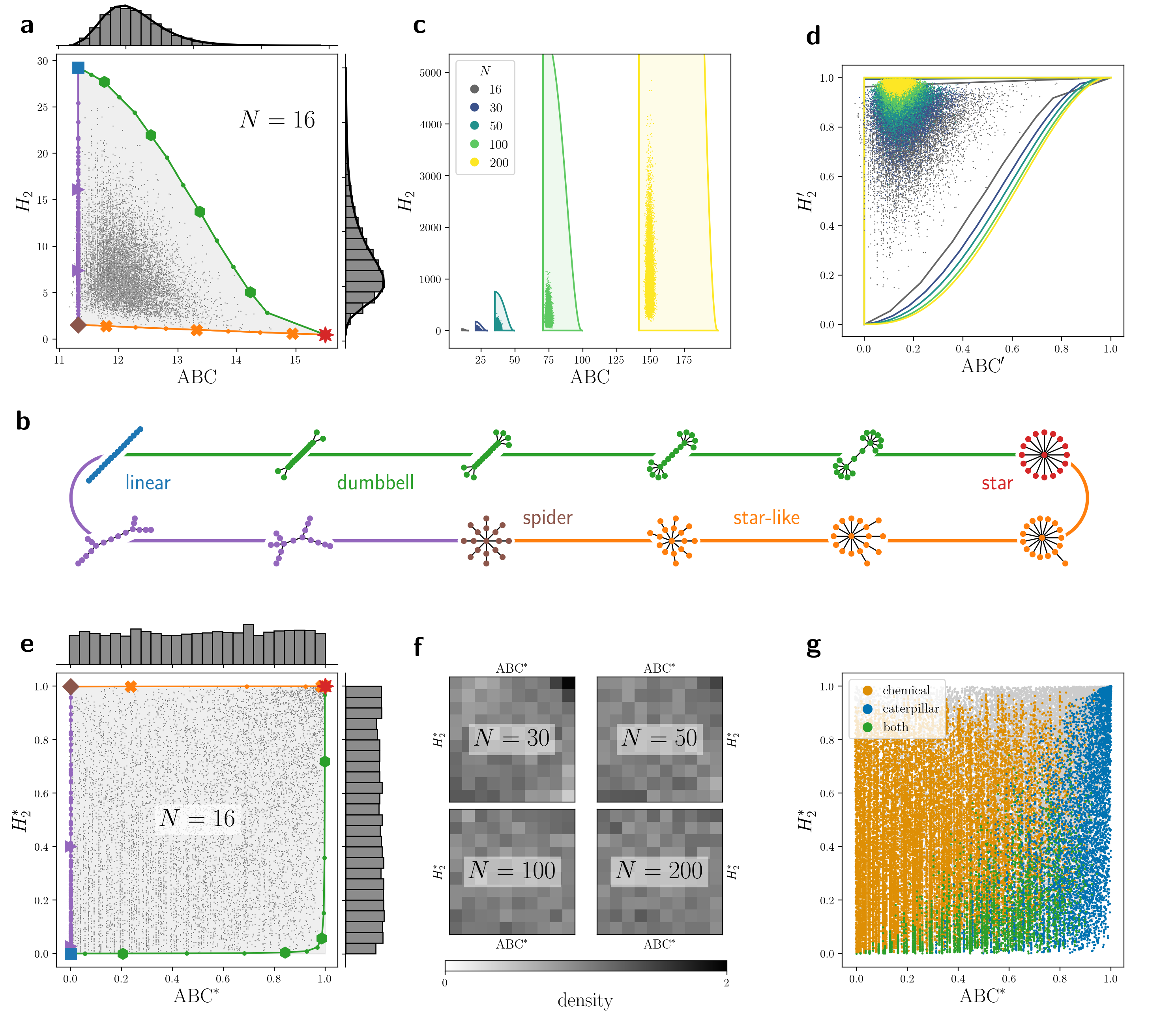}
    \caption{Phase space constructed from topological indices ABC and $H_2$. {\bf (a)} $(\ABC,H_2)$ space of $10^4$ random trees with $N=16$ edges. Distributions at the axes' boundaries show the probability distribution functions of individual indices. {\bf (b)} Different tree topologies that trace the boundary of the phase space, with large symbols in panels (a) and (d) corresponding to the shown tree structures. The trees at the boundary range from a linear tree, dumbbell trees, star tree, star-like trees, spider tree, and back to a linear tree; their construction is shown in Fig.~S5 in SI. {\bf (c)} $(\ABC,H_2)$ space for trees of four different sizes, where $10^4$ random trees are sampled for each $N$. {\bf (d)} $(\ABC',H_2')$ space of trivially normalized indices [Eq.~\eqref{eq:tip}] showing trees of four different sizes with $10^4$ random trees for each size (cf.\ panel (c)). {\bf (e)} $(\ABC^*, H_2^*)$ space of indices normalized through their cumulative distribution functions [Eq.~\eqref{eq:cdf}] for $10^4$ random trees with $N=16$. Distributions at the axes' boundaries show the probability distribution functions of individual indices. {\bf(f)} {Heatmaps of the probability density of trees in the} $(\ABC^*, H_2^*)$ space for $10^4$ random trees of size $N=30$, $50$, $100$, and $200$. {\bf (g)} $(\ABC^*, H_2^*)$ space of all non-isomorphic trees with $N=16$, showing separately chemical and caterpillar trees as well as their union.}
    \label{fig:2}
\end{figure*}

The phase space formed by the pair of indices $(\ABC,H_2)$ permits us to discriminate between tree topologies at {\em fixed} tree size $N$ (Fig.~\ref{fig:2}a) significantly better than if we would have used either of the two indices alone {or a different combination of two indices such as the commonly used Wiener and Randić indices $W$ and $R$ (Fig.~S6 in SI)}. We can furthermore trace the boundary of the $(\ABC,H_2)$ phase space with special cases of tree topologies, ranging from linear {and dumbbell trees} to star {trees}, star-like trees, and spider {trees} (Fig.~\ref{fig:2}b), which provides us with a basic understanding of topologies found in different regions of the phase space.

\subsection{Comparing topologies of trees of different sizes}

Topological indices, including $\ABC$ and $H_2$, show a very strong dependence on tree size $N$ (Fig.~S2 in SI). As Fig.~\ref{fig:2}c demonstrates, this prohibits a direct comparison of trees of different size. In order to do so, we need to normalize the indices to eliminate the dependence on $N$. The only normalization that has been used previously simply normalizes a topological index $\TI$ by its values for linear and star trees~\cite{balaban1979chemical,austel1983topological}, which are usually (but not always) also its extreme cases (Table~S1 in SI):
\begin{equation}
\TI'(T)=\frac{\TI(T)-\TI(\mathrm{linear})}{\TI(\mathrm{star}) -\TI(\mathrm{linear})}.
\label{eq:tip}
\end{equation}
Such normalization limits an index to the range $\TI'\in[0,1]$, where $\TI'(\mathrm{linear})=0$ and $\TI'(\mathrm{star})=1$. However, this turns out to be too simple for the purpose of comparison, as it progressively compresses the phase space as $N$ increases (Fig.~\ref{fig:2}d) and makes discrimination between different branching architectures difficult. The compression of the phase space is due to the mean values of a topological index and its values for linear and star trees all scaling differently with tree size (Table~S1 and Fig.~S2). In our case, this effect is much more pronounced for $H_2$ compared to $\ABC$.

We can significantly improve on the normalization of topological indices by focusing on their {\em cumulative distribution functions} ($\cdf$) instead of the indices themselves,
\begin{equation}
    \TI(T)\longrightarrow \cdf(\TI(T)).
\end{equation}
The cumulative distribution function is by definition limited to the range $\cdf(\TI(T))\in[0,1]$ irrespective of tree size $N$, with the boundaries of the range defined by the minimum and maximum value of the original topological index. Redefining a topological index through its cumulative distribution function also allows for a probabilistic interpretation of its value, as $\cdf(\TI(T))$ gives the probability of finding another tree between the minimum value $\TI_\mathrm{min}$ and the value of the topological index $\TI(T)$.

To determine the $\cdf(\TI)$ of a given topological index, we need to first obtain its probability density function, $\pdf(\TI)$. This is not a trivial endeavour since probability density functions for topological indices are not known. We therefore tested different continuous parametric distributions and chose the best-fitting one for each topological index (see Appendix and Fig.~S3 in SI). The $\pdf(\ABC)$ is best described by a log-normal distribution [Eq.~\eqref{eq:lnd}], while {the} Moyal distribution [Eq.~\eqref{eq:moy}] is the best choice for the $\pdf(H_2)$. The $N$-dependence of the fit parameters for both distributions is well-described by a simple power law and allows us to use the fitted $\pdf(\TI)$ to immediately determine the cumulative distribution function $\cdf(\TI)$ for an arbitrary tree size $N$.

The tree topologies assuming the extreme values of $\cdf(\TI(T))$ correspond to the tree topologies that assume the extreme values of the original index $\TI$. If we want to define a normalized index $\TI^*$ so that its values for the linear and the star tree are {\em fixed} to be $\TI^*(\mathrm{linear})=0$ and $\TI^*(\mathrm{star})=1$, we have to normalize the cumulative distribution function accordingly. We thus define a normalized topological index $\TI^*$ as:
\begin{equation}
   \TI^*(T) = \frac{\cdf(\TI(T)) - \cdf(\TI({\text{linear}}))}{\cdf(\TI({\text{star}})) - \cdf(\TI({\text{linear}}))},
   \label{eq:cdf}
\end{equation}
The probabilistic interpretation of the normalized index $\TI^*$ now refers to the probability of finding another tree $T$ between the values of $\TI(\mathrm{linear})$ and $\TI(T)$. Because $\ABC$ is not minimal for linear tree ($\ABC_\mathrm{min}\neq \ABC(\mathrm{linear)}$, see Table~S1), a vanishing fraction of trees will obtain $\ABC^* < 0$ (Fig.~S7 in SI); for these, the probabilistic interpretation of $\ABC^*$ does not hold.

\subsection{Properties of \texorpdfstring{$(\ABC^*,H_2^*)$}{TEXT} space}

\begin{figure*}[!htp]
    \centering
    \includegraphics[width=\linewidth]{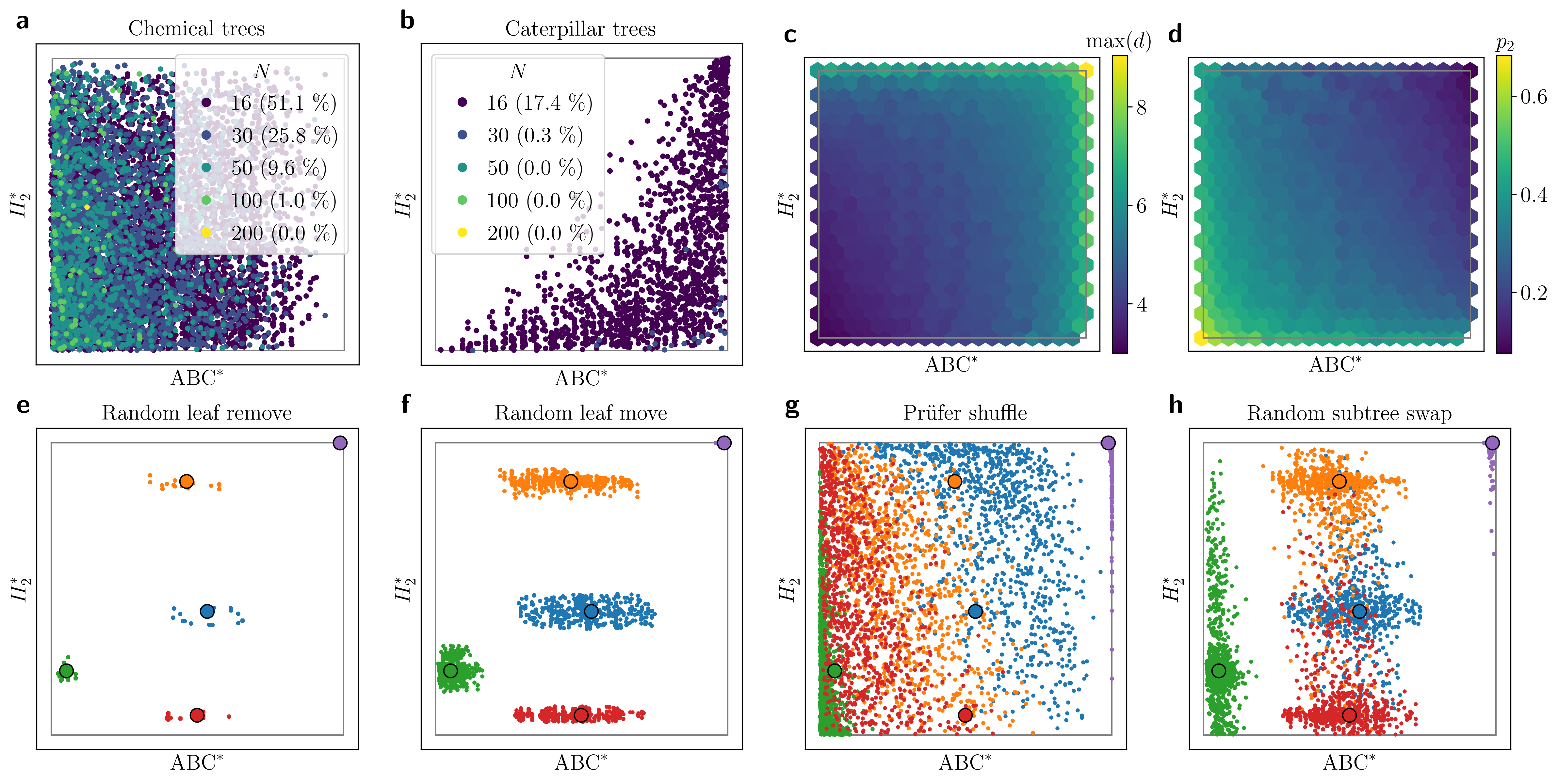}
    \caption{Characterization of the $(\textrm{ABC}^*,H_2^*)$ space. Changes in the proportion of {\bf (a)} chemical and {\bf (b)} caterpillar trees in the phase space with increasing tree size $N$, estimated from samples of $10^4$ random trees. Heatmaps of {\bf (c)} maximum node degree and {\bf (d)} fraction of degree $2$ nodes for trees with $N=16$. {\bf (e)}--{\bf (h)}: Changes in the phase space for trees with $N=50$ upon performing four different types of tree transformations $10^3$ times on five randomly chosen trees (black circles): random leaf move, random leaf remove, Prüfer shuffle, and random subtree swap. See the main text and Appendix for details on these transformations.}
    \label{fig:2.5}
\end{figure*}

We can now use the normalization via cumulative distribution functions to create a space of normalized indices $(\ABC^*,H_2^*)$ which leads to a significantly more homogeneous phase space and enables us to compare topologies of trees of arbitrary size (panels (e) and (f) of Fig.~\ref{fig:2}). The $(\ABC^*,H_2^*)$ space also allows us to interpret the relationships between different tree topologies at its boundary (Fig.~\ref{fig:2}e) as well as inside it. For instance, chemical trees, which have a maximum branching degree of $4$, and caterpillar trees, whose nodes all lie within distance $1$ of a central path, occupy distinct parts of the phase space (Fig.~\ref{fig:2}g). The part of the phase space that chemical and caterpillar trees occupy furthermore shrinks with increasing tree size (panels (a) and (b) of Fig.~\ref{fig:2.5}) since their overall proportion compared to all trees decreases with increasing $N$. More generally, the position and proportion of different types of trees in the $(\ABC^*,H_2^*)$ space is broadly determined by their maximum node degree and number of degree $2$ nodes. Specifically, the maximum node degree of a tree in the phase space increases as we go from a linear tree at $(0,0)$ to a star tree at $(1,1)$ (Fig.~\ref{fig:2.5}c); at the same time, the number of degree $2$ nodes diminishes (Fig.~\ref{fig:2.5}d).

We can also use the pair of normalized indices $(\ABC^*,H_2^*)$ to examine how different transformations of trees reposition their topology in the phase space. If we either move or remove a random leaf node, this repositions the new tree topology along the $\ABC$ axis but leaves its position along the $H_2$ axis fairly unchanged (panels (e) and (f) of Fig.~\ref{fig:2.5}). Transformations that affect larger parts of the tree topology---such as the Pr\"ufer shuffle, which retains the node degree distribution of a tree but rearranges the connections between nodes, and random subtree swap, where an edge is randomly deleted and a new one randomly added---change the position of tree topology in the phase space more drastically (panels (g) and (h) of Fig.~\ref{fig:2.5}).

Understanding where different tree topologies and their transformations lie in the $(\ABC^*,H_2^*)$ space enables us to discriminate between different architectures of branched macromolecules. In the following two sections, we demonstrate the usefulness of the phase space of normalized topological indices on two practical examples. We first focus on the comparison of different methods of mapping highly-branched RNA structures to trees and of differences in their topology. Next, we compare different methods of coarse-graining trees that can be applied to general branched macromolecules in order to reduce their topology for the purposes of numerical or theoretical modelling.

\begin{figure*}[!htp]
    \centering
    \includegraphics[width=\linewidth]{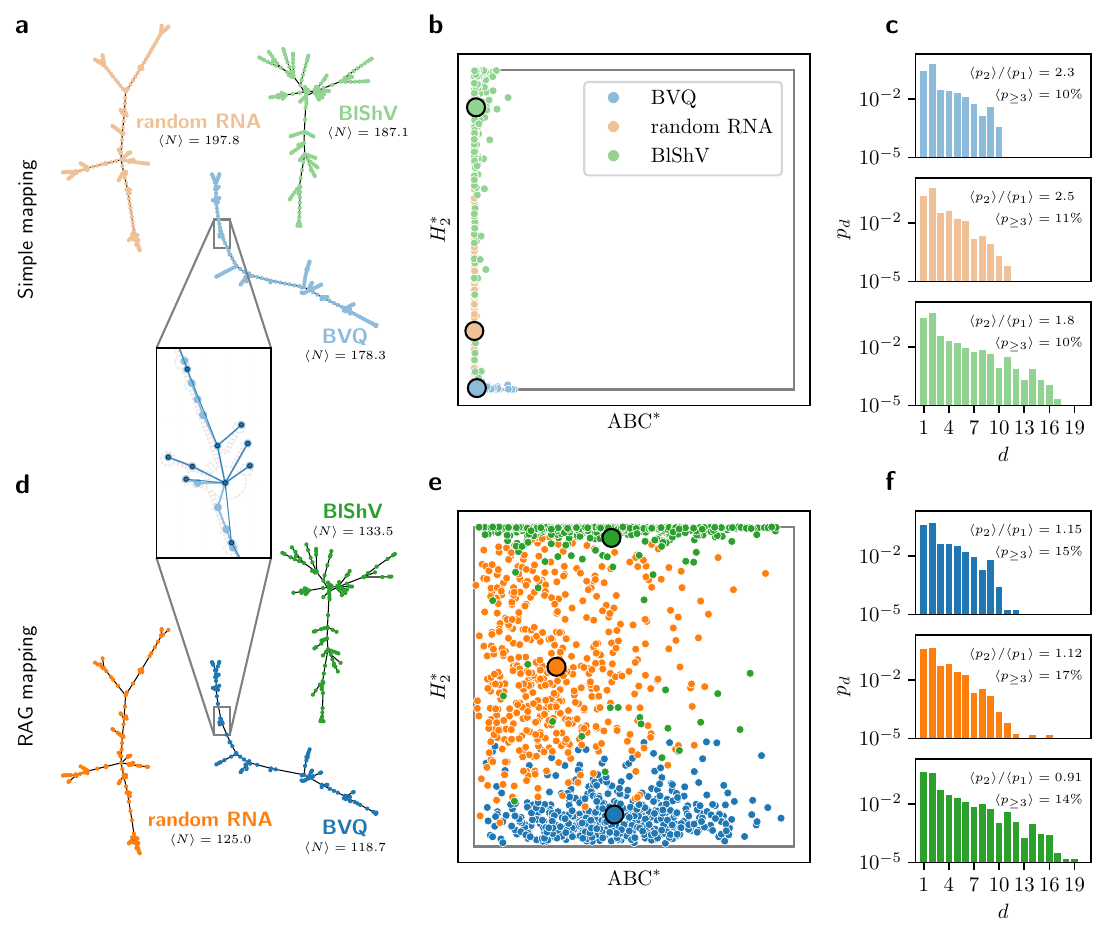}
    \caption{Comparison of two methods of mapping RNA secondary structures to mathematical trees, {\bf (a)}--{\bf (c)} simple mapping and {\bf (d)}--{\bf (f)} RAG mapping (see main text for details). Panels (a) and (d) show example secondary structures of three different RNA sequences: non-compact genome of beet virus Q (BVQ), compact genome of blueberry shock virus (BlShV), and random RNA sequence of uniform nucleotide composition. The trees resulting from simple (panel (a)) and RAG (panel (d)) mapping differ in their average size $\langle N\rangle$. Boxed region between the two panels illustrates the differences in the two mapping methods on a part of (the same) RNA structure to a tree (cf.\ also Fig.~S8). Panels (b) and (e) show the positions of $500$ trees in the $(\ABC^*,H_2^*)$ space obtained from RNA secondary structures by simple and RAG mapping, respectively. {Each point corresponds to a tree mapped from a secondary structure sampled from the thermal ensemble (Methods and Appendix). Points highlighted by black edges show the ensemble averages of the topological indices for each of the three RNA sequences.} Panels (c) and (f) show the node degree distributions of the $500$ trees of each RNA sequence obtained from simple and RAG mapping, respectively, where $p_d$ denotes the fraction of nodes with degree $d$ in a tree. Averages of different quantities $\langle \cdot \rangle$ are always performed over $500$ trees.}
    \label{fig:3}
\end{figure*}

\section{Example: different tree representations of RNA structure}

Base-pairing between the four constituent nucleotides of RNA leads to a branched secondary structure~\cite{vaupotivc2023scaling,vaupotivc2023viral} with numerous high-degree branching nodes ($d(i)>4$)~\cite{wiedemann2022rnaloops}, making RNA unique among known branched macromolecules. This often carries functional significance: For instance, highly-branched RNA genomes of some viruses are compact and consequently more easily encapsidated during virion formation than less branched ones~\cite{yoffe2008predicting,tubiana2015synonymous}. Graphs present a natural way to coarse-grain RNA structure and analyze its branching topology~\cite{Schlick2018}. However, the choice of the mapping of the base-pair pattern onto a mathematical tree is not straightforward and several options exist, designed with different purposes in mind~\cite{kim2013network,shapiro1988algorithm}. Perhaps the simplest way to map RNA secondary structure to a tree is by mapping double-stranded regions (base pairs) to edges, while single-stranded regions (unpaired nucleotides) are mapped to nodes connecting the edges~\cite{vaupotivc2023scaling,vaupotivc2023viral}. An alternative way, used in the RNA-as-graph (RAG) framework~\cite{Schlick2018}, omits from nodes those unpaired regions which consist of a single unpaired nucleotide or a single non-complementary pair of nucleotides, and does not consider lonely base pairs as edges. (For details on the two approaches, see Appendix.)

While the simple and RAG-based methods of representing RNA as a tree appear very similar, they lead to quite different trees due to the properties of typical RNA secondary structure folds (Fig.~S8 in SI). Furthermore, these trees will have different size, making it complicated to compare them using non-normalized topological indices. The phase space of normalized topological indices $(\ABC^*,H_2^*)$ thus provides an excellent framework for the comparison of the two mapping methods (Fig.~\ref{fig:3}). As an example, we focus on three distinct RNA sequences and their secondary structures. We compare the RNA genomes of beet virus Q (BVQ) and blueberry shock virus (BlShV), which were previously found to be among the least and most compact viral RNA genomes, respectively, using maximum ladder distance (MLD; corresponding to tree diameter) as a measure~\cite{vaupotivc2023viral}. Additionally, we include a random RNA sequence of uniform nucleotide composition and similar length for comparison.

Secondary structures of the three RNA sequences are sampled from their thermal ensembles ({Methods and Appendix}) and then mapped to trees using either the simple or the RAG method. While the RAG method (Fig.~\ref{fig:3}d) results in trees on average two thirds the size of those obtained using the simple method (Fig.~\ref{fig:3}a), the $(\ABC^*,H_2^*)$ space allows us to compare the RNA topologies resulting from the two mapping methods (panels (b) and (e) of Fig.~\ref{fig:3}). For both methods we observe that the three RNA sequences are separated in the phase space along the $H_2^*$ axis. The compact BlShV genome occupies large $H_2^*$ values, the non-compact BVQ genome occupies small $H_2^*$ values, and the random RNA sequence is located in-between the two. The $\ABC^*$ index, on the other hand, shows the differences between the two mapping methods. For the simple method, most trees acquire values of $\ABC^*$ close to zero, {showing little fluctuation between the structures sampled from the thermal ensemble}. For the RAG method, however, the $\ABC^*$ values of both RNA genomes are distributed along the entire range of values, {implying larger thermal fluctuations between the structures}; $\ABC^*$ values of the random RNA, while non-zero, remain small and in this way differ from both biological RNAs. These differences can be traced to the proportion of degree $2$ nodes produced by the two mapping methods, as the simple method (Fig.~\ref{fig:3}c) results in approximately twice as many degree $2$ nodes than the RAG method (Fig.~\ref{fig:3}f). As we have already observed previously, the amount of degree $2$ nodes is an important factor that determines the position of a tree topology in the phase space (Fig.~\ref{fig:2.5}).

The two normalized indices, $\ABC^*$ and in particular $H_2^*$, are thus able to discriminate between different types of RNA sequences and their structures, regardless of how they are mapped to a tree---and even when their branching properties, such as the distribution of branching degrees, appear similar. At the same time, the two indices are able to elucidate the main differences in the two mapping approaches we have used to obtain RNA trees, the simple and the RAG method. The RAG method has a higher resolution in the $(\ABC^*,H_2^*)$ space compared to the simple one and can thus discriminate better between RNA trees.

\section{Example: Coarse-graining the topologies of random trees}

Methods of mapping RNA secondary structures to trees can also be seen as different levels of a coarse-grained representation of RNA structure. More generally, the question of how to coarse-grain the structure of branched macromolecules is important both from the perspective of the theory of branched polymers as well as from the perspective of large-scale molecular dynamics simulations. Mean-field theoretical models of RNA branching, for instance, show a strong dependence of RNA encapsidation on the assumed (coarse-grained) branching properties of RNA~\cite{farrell2023role,erdemci2016effects,erdemci2014rna,marichal2021relationships}. In simulations of polymers, the extent to which coarse-grained models~\cite{gartner2019modeling} capture polymer topology and its related physics is an important consideration that can drastically affect predicted dynamics and thermodynamic properties~\cite{dhamankar2021chemically,Xia2019}.

\begin{figure*}[!htp]
    \centering
    \includegraphics[width=\linewidth]{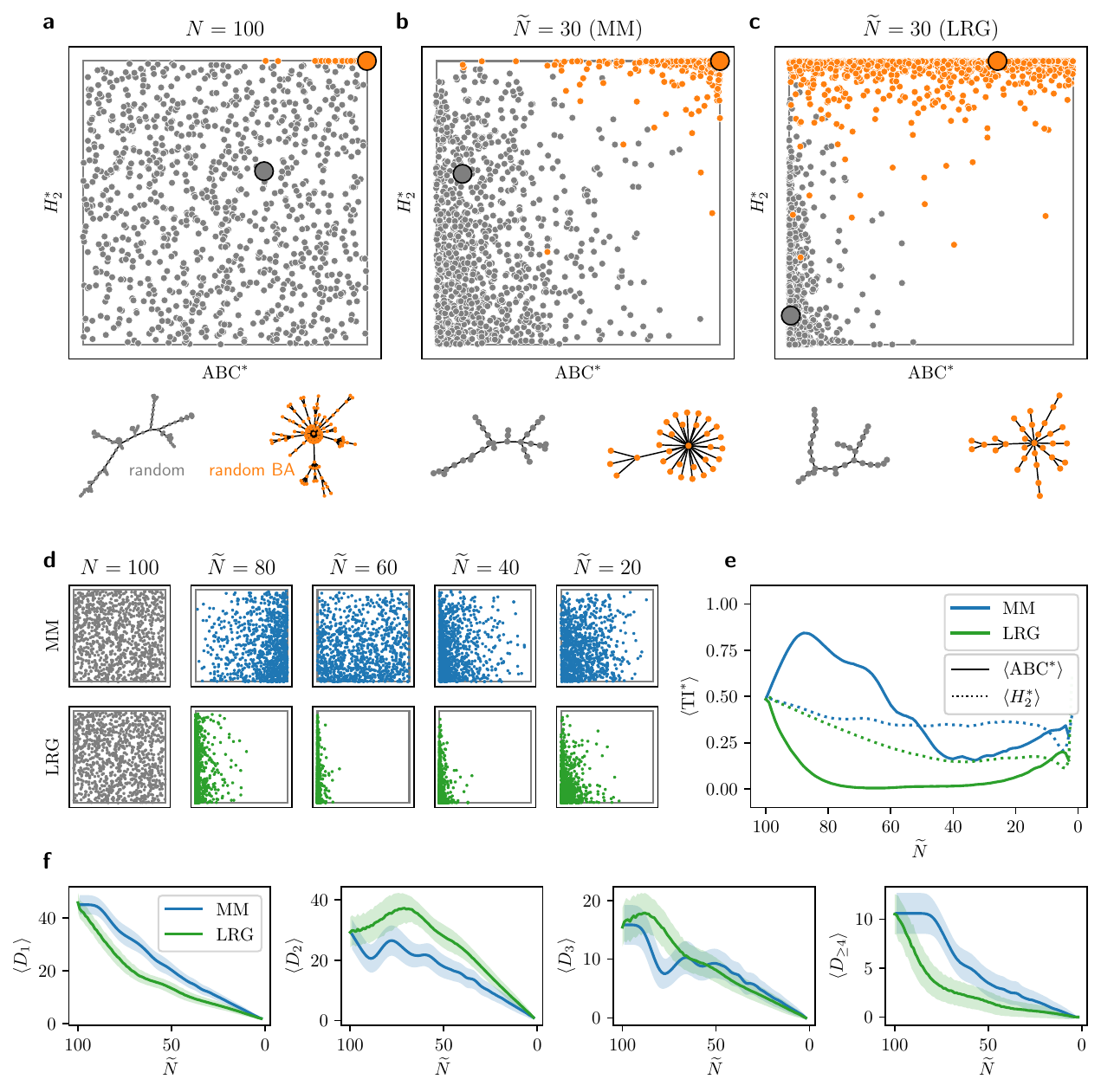}
    \caption{Two methods of coarse-graining trees and their effects on branching topology. {\bf (a)} $(\ABC^*,H_2^*)$ space of $10^3$ random trees and $10^3$ random BA trees with $N=100$ and the phase space of these trees after undergoing {\bf (b)} MM coarse-graining and {\bf (c)} LRG coarse-graining reducing their size to $\widetilde{N}=30$. Example tree structures below panel (a) and their coarse-grained versions below panels (b) and (c) correspond to the points highlighted by black circles in the phase space. {\bf (d)} Phase space flow of $10^3$ random trees upon their coarse-graining from $N=100$ to progressively smaller target sizes $\widetilde{N}$ using either MM (top row) or LRG (bottom row). {{\bf (e)} Changes in the average values of normalized topological indices $\langle\ABC^*\rangle$ and $\langle H_2^*\rangle$ and {\bf (f)} changes in the average number of nodes with degree $i$, $\langle D_i\rangle$, with the size of coarse-grained trees $\widetilde{N}$ using MM and LRG. All averages are performed over $10^3$ random trees. Shaded regions indicate the range of one standard deviation from the mean.}}
    \label{fig:4}
\end{figure*}

Any coarse-graining by definition leads to a loss of information, and we can use the $(\ABC^*,H_2^*)$ space to examine in detail the consequences of coarse-graining for the reduced topology of branched macromolecules. For simplicity, we will start already from mathematical trees and explore the effect of different strategies to reduce their size and coarse-grain them into smaller trees. We focus on two examples of common graph coarse-graining methods, modularity-maximization (MM) coarse-graining and Laplacian renormalization group (LRG) coarse-graining (see Appendix). Starting with trees of size $N=100$, we coarse-grain them to different target tree sizes $\widetilde{N}$ and study how their position in the $(\ABC^*,H_2^*)$ space changes (Fig.~\ref{fig:4}). We do this both for general (random) tree topologies as well as for scale-free Barabási-Albert (BA) trees.

First, we see that scale-free BA trees occupy only a very small part of the $(\ABC^*,H_2^*)$ space with high $H_2^*$ and $\ABC^*$ (Fig.~\ref{fig:4}a), which we have previously identified as corresponding to very star-like trees (Fig.~\ref{fig:2}e). Random trees are, on the other hand, distributed across the entire phase space of normalized indices, as we have already seen in Fig.~\ref{fig:2}. When both types of trees with initial size $N=100$ are coarse-grained to $30\%$ of their size (i.e., $\widetilde{N}=30$), the two coarse-graining methods produce quite different results. MM retains the approximate clustering of BA trees in the upper right corner of the phase space while compressing random trees to the left side of the space (Fig.~\ref{fig:4}b). LRG is less effective in preserving the topological properties as captured by the phase space (Fig.~\ref{fig:4}c): while they preserve the values of $H_2^*$ of BA trees, their values of $\ABC^*$ get spread out; random trees are squeezed to very low values of $\ABC^*$. Example trees in Fig.~\ref{fig:4} provide a visual intuition of the differences between the two methods: On the one hand, MM better preserves the global tree topology because it is optimised in terms of connectivity within a cluster. On the other hand, LRG is better at conserving the fine details of the branching topology because it optimizes the clusters in terms of ``diffusive'' distance.

We next progressively coarse-grain random trees towards smaller sizes $\widetilde{N}$ and monitor their flow in the $(\ABC^*,H_2^*)$ space (panels (d) and (e) of Fig.~\ref{fig:4}). Again, MM better preserves the properties of random trees in the phase space compared to LRG, which always pushes trees towards small values of $\ABC^*$---i.e., towards more linear trees. {This behaviour is even more obvious if we examine trajectories of individual trees during the coarse-graining process. Figure~S9 in SI shows examples of four trees chosen from Fig.~\ref{fig:4}d, and we observe that LRG reduces the $\ABC^*$ value of trees to almost zero already at $80\%$ coarse-graining. This is not the case for MM, where the $\ABC^*$ values typically first increase and only gradually start moving towards smaller values as the trees are coarse-grained further. While $\ABC^*$ changes non-monotonically with MM coarse-graining, $H_2^*$ exhibits a plateau at coarse-graining levels below $\lesssim70\%$ (Fig.~\ref{fig:4}e).} When we monitor the same flow while coarse-graining scale-free BA trees (Fig.~S10 in SI), their position in the $(\ABC^*,H_2^*)$ space is preserved well until the coarse-grained tree size $\widetilde{N}$ becomes very small; the only exception is again the effect of LRG on $\ABC^*$, as {the latter again decreases quickly}. {The different effects of MM and LRG on the topological indices of trees---and in particularly on $\ABC^*$---can be explained by taking a look into the distributions of node degrees in the coarse-grained trees (Fig.~\ref{fig:4}f). LRG initially increases the numbers of degree 2 and 3 nodes, and as we have already seen with RNA in Sec.~III, a high proportion of degree $2$ nodes leads to very low values of $\ABC^*$. In contrast, MM at no point during the coarse-graining increases the number of nodes of any degree, but appears instead to initially preserve the number of high degree nodes in the coarse-grained trees.}

Any coarse-graining applied to the structure of a branched macromolecule will necessarily change its underlying topology. Our framework---using the $(\ABC^*,H_2^*)$ space or even another combination of indices of interest to monitor the effects of coarse-graining methods on branching topology---provides a clear way to choose both the best coarse-graining strategy as well as the {\em extent} of coarse-graining most suited to the system and properties under examination. {While the choice of the best coarse-graining method strongly depends on its purpose, if the coarse-graining is meant to capture the essential features of tree topologies and retain the differences between them, a good method should {\em maintain the position of the structures} in the $(\ABC^*,H_2^*)$ space as much as possible during coarse-graining. Even though this calls for an in-depth study of different coarse-graining methods, our results (Fig.~\ref{fig:4}) indicate that between MM and LRG, this criterion strongly favours MM, with coarse-graining to $\sim50\%$ of the initial tree size yielding the most consistent results.}

\section{Discussion and conclusions}

In this work, we introduced a novel normalization of topological indices using the estimates of their cumulative distribution functions. We used this normalization to create a two-dimensional phase space based on atom bond connectivity index $\ABC^*$ and second-order network coherence $H_2^*$, which enables discrimination between branching architectures of macromolecules of different types and sizes. We demonstrated the usefulness of this phase space by exploring the topologies of different RNA structures mapped to trees using two different methods and by comparing the effects of two distinct tree coarse-graining methods on the resulting coarse-grained topologies.

Since topological indices are defined on graphs, our approach can also be---in principle---extended from trees to graphs. This extension is, however, not trivial, since the space of all possible graph topologies is significantly larger compared to trees and the probability density function of topological indices consequently more difficult to estimate. In Fig~S12 in SI we show how the phase space of {\em non-normalized} indices $(\ABC,H_2)$ looks like for all graphs having up to three cycles. The same two distributions were fitted as for trees [Eqs.~\eqref{eq:lnd} and~\eqref{eq:moy}], and while the $\ABC$ seems to still follow the log-normal distribution, a better distribution should be found for $H_2$. With suitable models for the probability distribution functions, however, the same normalization as we have applied to trees [Eq.~\eqref{eq:cdf}] should allow one to obtain a uniform phase space of normalized topological indices also for graphs containing several cycles.

We have shown that the phase space of the two normalized topological indices $(\ABC^*,H_2^*)$ allows for a robust comparison of molecules of different sizes---stemming either from differences in their molecular weight or their mapping onto mathematical trees. Our choice of topological indices is important, as not all combinations of topological indices lead to such a well-defined and interpretable phase space. Figure~S13 in SI shows an example of another phase space, defined through the commonly-used Wiener index $W$ and Randi\'{c} index $R$, $(W^*,R^*)$ (see SI for their definitions). In this case, the normalization using Eq.~\eqref{eq:cdf} leads to a far less homogeneous phase space compared to $(\ABC^*,H_2^*)$ space, {which comes in addition to the reduced discrimination power of these two indices (Fig.~S6)}. {Approaches using a linear combination of several indices, extracted via principal component analysis (PCA), are also commonly used in QSPR studies~\cite{basak1987topological,perdih2000topological}. Such approaches, however, significantly impact both the interpretability of the phase space spanned by the principal components as well as the determination of the probability density functions needed for normalization. Similarity between graph-like objects can also be studied using machine learning-based approaches, where graph vertices are mapped to a low-dimensional vector space based on local characteristics and the embedding is then optimized by reconstructing the graph from it and checking how much it deviates from the original~\cite{hamilton2017representation,coley2017convolutional,li2022deep,fung2021benchmarking}. Such methods can be used to derive similarity measures based on the distance between two embeddings~\cite{li2019graph}; whether or not they are also applicable in lieu of normalized topological indices---in particular when it comes to their interpretability and the ability to compare structures of different size---requires a carefully set-up comparison and remains a topic for future work~\cite{sabando2022using}.}

Using the phase space of normalized topological indices can be applied very broadly to discuss and compare different methods to coarse-grain the structure of large branching macromolecules and determine the best level and method of coarse-graining that retains the desired properties of their branching topology. This approach can be particularly useful for modelling long RNA sequences in biologically-informed RNA studies, e.g., in those that seek to single out conserved topological properties in viral genomes, or for long non-coding RNAs. It might also offer a valuable tool in the future to describe different architectures of lignin, a macromolecule characterized by large structural diversity. Lastly, an extension of our approach from trees to graphs would prove indispensable for the characterization of not not only more complex models of RNA folding but also of protein interaction networks~\cite{lambrughi2021pyinteraph} and gene interactions in Hi-C maps of the genome.

\section*{Supplementary Material}

See the supplementary material for definitions of topological indices and connections between them; details on normalization of topological indices through their cumulative distribution functions; details on characterization of the $(\ABC^*,H_2^*)$ space; methods of mapping RNA secondary structure to trees; methods of coarse-graining trees; extending the phase space of topological indices to graphs; and details on the $(W^*,R^*)$ space.

\section*{Acknowledgments}

D.V.\ and A.B.\ acknowledge support by Slovenian Research Agency (ARRS) under contracts no.\ P1-0055 and no.\ J1-60002. J.M.\ and L.T.\ acknowledge support by ICSC---Centro Nazionale di Ricerca in High Performance Computing, Big Data and Quantum Computing, funded by European Union---NextGenerationEU. J.M.\ acknowledges support by the European Union under NextGenerationEU grant PRIN 2022 PNRR Prot P2022TX4FE. J.M.\ and L.T.\ also thank the University of Trento for the strategic project ``AIACE'' that enabled this collaboration.

\section*{Author declarations}

\subsection*{Conflict of interest}

The authors declare no competing interest.

\subsection*{Author contributions}

{\bf Domen Vaupotič}: Conceptualization (equal); Formal analysis (lead); Investigation (equal); Methodology (equal); Visualization (lead); Writing – review \& editing (equal). {\bf Jules Morand}: Conceptualization (equal); Investigation (equal); Methodology (equal); Writing – review \& editing (equal). {\bf Luca Tubiana:} Conceptualization (equal); Investigation (equal); Methodology (equal); Writing – review \& editing (equal). {\bf Anže Božič}: Conceptualization (equal); Investigation (equal); Methodology (equal); Supervision (lead); Writing – original draft (lead); Writing – review \& editing (equal).

\section*{Data and code availability}

Data supporting the findings of this study are available within the paper and supporting information. Code related to this work is publicly available at \url{https://github.com/ljRNA/topological-indices}.

\appendix

\section*{Appendix: Extended Methods}

\subsection*{Tree generation}

To generate different types of trees, we use functions from NetworkX~3.2 package~\cite{hagberg2008exploring}. Specifically, we use \texttt{nonisomorphic\textunderscore trees} to generate sets of {\em all} non-isomorphic trees, \texttt{random\textunderscore unlabeled\textunderscore tree} to generate a uniformly sampled random set of unlabelled trees, and \texttt{barabasi\textunderscore albert\textunderscore graph} to generate random Barabási--Albert trees; all of them at fixed $N$.

\subsection*{Calculation of topological indices}

Path-based indices were calculated using Floyd-Warshall algorithm in NetworkX~3.2~\cite{hagberg2008exploring}. Node automorphism orbits, required for the index $I$, were calculated using nauty~2.8.8~\cite{mckay2014practical}.

\subsection*{Distribution functions of topological indices}

To the best of our knowledge, no distribution of any topological index for trees of a given size is known. We have thus tested 101 different continuous parametric distributions to find the best approximations to the empirical distributions of topological indices. Our criteria included goodness-of-fit (as measured by Bayesian information criterion, BIC) at various $N$, low complexity (small number of parameters), as well as smooth scaling of extracted parameters with $N$.

For the two indices we focus on, ABC and $H_2$, we find that the best choices describing their probability density functions are the log-normal distribution
\begin{equation}
    \label{eq:lnd}
        f(x; s, a, b) = \frac{1}{\sqrt{2\pi}\, (x-a)s} \exp\left(-\frac{\left[\log(\frac{x-a}{b})\right]^2}{2s^2}\right)
\end{equation}
for the ABC index and the Moyal distribution
\begin{equation}
    \label{eq:moy}
        f(x; a, b) = \frac{1}{\sqrt{2\pi}\,b} \exp\left(-\frac{x-a}{2b} - \frac12\exp\left(-\frac{x-a}{b}\right)\right)
\end{equation}
for the $H_2$ index. Therefore, the log-normal distribution is fitted using three parameters $p$ ($s$, $a$, and $b$), and the Moyal distribution using two parameters $p$ ($a$ and $b$).
 
{The distribution parameters $p$ are first estimated with maximum likelihood estimation on sets of $10^4$ random trees at various $N$ ranging from $N=10$ to $N=300$ (Fig.~S3 in SI). Sampling the values of topological indices on $10^4$ random trees is sufficient to obtain a good approximation to the distributions of the indices over the space of all (non-isomorphic) trees (Fig.~S4 in SI).} Afterwards, we fit the dependency of each distribution parameter $p$ on tree size $N$ with a power-law model of the form $p = \alpha N^\beta$ (Fig.~S3). The power-law estimates of the distribution parameters are then used to obtain the cumulative distribution function of a given topological index for a tree of arbitrary size $N$.

\subsection*{Tree transformations}

We employ four different tree transformations to study where the transformed trees lie in the phase space of normalized topological indices. Namely, we use:
\begin{itemize}
    \item Random leaf remove: A randomly selected leaf (degree 1 node) is removed; 
    \item Random leaf move: A randomly selected leaf (degree 1 node) is removed and a new leaf is added to a randomly selected node in a tree;
    \item Prüfer shuffle: The Prüfer sequence of a tree is randomly shuffled, leading to a tree with a same distribution of node degrees $d(i)$ but a different topology (see, e.g., Refs.~\cite{singaram2016prüfer,vaupotivc2023scaling});
    \item Random subtree swap: A randomly selected edge is removed, resulting in two disconnected trees; a new edge is then added between two randomly selected nodes of the two trees to form a new tree.
\end{itemize}

\subsection*{RNA sequences and prediction of their secondary structure}

We consider three distinct examples of RNA sequences: two genomes of single-stranded RNA viruses and a random RNA sequence with uniform nucleotide composition~\cite{vaupotivc2023viral}. The two viral RNA sequences were obtained from the GenBank database~\cite{genbank}: RNA 2 of blueberry shock virus (KF031038) with a length of 2603~nt, and RNA 3 of beet virus Q (AJ223598) with a length of 2529~nt. The length of the random RNA sequence is 2600~nt. To obtain the secondary structures of the three RNA sequences, we use ViennaRNA software (v2.6.4)~\cite{Lorenz2011}. Specifically, for each sequence we generate 500 secondary structure folds sampled from the thermal ensemble ($T=37\;{}^\circ\mathrm{C}$) using the \texttt{RNAsubopt} routine with default settings~\cite{vaupotivc2023scaling}. The secondary structure prediction does not include pseudoknots, which is a typical simplification that also allows us to map the RNA secondary structure to a tree.

\subsection*{Mapping RNA to a tree}

We employ two different methods of mapping an RNA secondary structure to a tree, where the main difference is in the way certain structural elements---namely, internal loops, bulges and lone base pairs---are treated. The first, simple method is based on two rules only~\cite{vaupotivc2023viral,vaupotivc2023scaling}:
\begin{enumerate}
    \item Each bulge, hairpin loop, internal loop, multi loop, and external loop is considered a node;
    \item Each double-stranded RNA stem is represented as an edge.
\end{enumerate}
The second method is based on the RAG approach~\cite{Schlick2018,Izzo2011} and uses a more complicated set of rules:
\begin{enumerate}
    \item Each bulge, hairpin loop, and internal loop is considered a node when there is more than one unmatched nucleotide or non-complementary base pair;
    \item Each multiloop and external loop is a node;
    \item Each double-stranded RNA stem with more than one complementary base pair is represented as an edge.
\end{enumerate}
An example illustrating both approaches is shown in Fig.~S8. We also note that while a mapping of RNA to a tree can assign weights to the edges (usually corresponding to the number of basepairs contained in them), we focus here only on the topology and thus consider each tree as unweighted.

\subsection*{Coarse-graining trees}

We use two different tree coarse-graining procedures, one based on modularity maximization (MM)~\cite{fortunato2010community} and the other on Laplacian renormalization group (LRG)~\cite{Villegas2023}. In MM, one finds sets of nodes called clusters that tend to maximize the number of edges inside them; these become new nodes in the coarse-grained tree. We perform MM with the Clauset-Newman-Moore greedy modularity maximization algorithm, implemented in NetworkX~3.2~\cite{hagberg2008exploring}. In LRG, nodes are clustered via a network propagator constructed from the graph Laplacian matrix and correspond to the ``communicability'' or diffusion trajectories in a graph. We implement LRG based on the work by \citet{Villegas2023}. The coarse-graining parameter in both MM and LRG methods is the target size (number of edges) of the coarse-grained tree $\widetilde{N}$. We note that while $\widetilde{N}$ can always be reached exactly using the MM method, using the LRG method leads to a (narrow) distribution of values centred around the target $\widetilde{N}$ (Fig.~S11 in SI).

\bibliography{references}


\newif\ifarXiv
\arXivtrue 

\ifarXiv
    \foreach \x in {1,...,\numbersupplementpages}
    {
        \clearpage
        \includepdf[pages={\x,{}}]{\supplementfilename}
    }
\fi

\end{document}